\pdfoutput=1

\documentclass[sigconf]{acmart}

\AtBeginDocument{%
  }

\setcopyright{acmlicensed}
\copyrightyear{2018}
\acmYear{2018}
\acmDOI{XXXXXXX.XXXXXXX}

\acmConference[Conference acronym 'XX]{Make sure to enter the correct
  conference title from your rights confirmation emai}{June 03--05,
  2018}{Woodstock, NY}
\acmISBN{978-1-4503-XXXX-X/18/06}

\usepackage{algorithm}
\usepackage{algorithmic}
\usepackage{array}
\usepackage{enumitem}
\usepackage{graphics}
\usepackage{graphicx}
\usepackage{makecell}
\usepackage{multirow}
\usepackage{subfigure}

\begin{document}

\title{Retrievable Domain-Sensitive Feature Memory for Multi-Domain~Recommendation}


\author{
Yuang Zhao$^{2*}$, 
Zhaocheng Du$^{1*}$\authornote{Equal contribution.},
Qinglin Jia$^{1}$,
Linxuan Zhang$^{2\dagger}$\authornote{Corresponding authors.},
Zhenhua Dong$^{1\dagger}$,
Ruiming Tang$^{1}$}
\renewcommand{\authors}{
Yuang Zhao,
Zhaocheng Du,
Qinglin Jia,
Linxuan Zhang,
Zhenhua Dong,
Ruiming Tang}
\affiliation{%
  \institution{
  $^1$Noah's Ark Lab,
  Huawei \quad 
  $^2$ Tsinghua University
  }
  \country{}
}
\email{
zhaoya22@mails.tsinghua.edu.cn,
lxzhang@tsinghua.edu.cn}
\email{{
zhaochengdu,
jiaqinglin2,
dongzhenhua,
tangruiming}@huawei.com}

\renewcommand{\shortauthors}{Zhao and Du et al.}

\begin{abstract}
With the increase in the business scale and number of domains in online advertising, multi-domain ad recommendation has become a mainstream solution in the industry. 
The core of multi-domain recommendation is effectively modeling the commonalities and distinctions among domains. 
Existing works are dedicated to designing model architectures for implicit multi-domain modeling while overlooking an in-depth investigation from a more fundamental perspective of feature distributions.
This paper focuses on features with significant differences across various domains in both distributions and effects on model predictions.
We refer to these features as domain-sensitive features, which serve as carriers of domain distinctions and are crucial for multi-domain modeling.
Experiments demonstrate that existing multi-domain modeling methods may neglect domain-sensitive features, indicating insufficient learning of domain distinctions. 
To avoid this neglect, we propose a domain-sensitive feature attribution method to identify features that best reflect domain distinctions from the feature set. 
Further, we design a memory architecture that extracts domain-specific information from domain-sensitive features for the model to retrieve and integrate, thereby enhancing the awareness of domain distinctions. 
Extensive offline and online experiments demonstrate the superiority of our method in capturing domain distinctions and improving multi-domain recommendation performance.
\end{abstract}




\maketitle

\section{Introduction}

Accurately predicting the probability of user feedback, such as click-through and conversion rate, is essential for precise ad recommendations~\cite{ctr_chapelle2014simple_ctr, ctr_richardson2007predicting_ctr, ctr_survey_chen2016deep_ctr}.
Ad recommendation generally involves multiple domains corresponding to various business scenarios~\cite{multidomain_zhu2017optimized_multidomain}.
The traditional practice is to train predicting models for different domains individually.
However, it faces challenges, including higher model maintenance costs and tail domain problems, as the business scale and number of domains increases~\cite{hmoe_li2020improving, sarnet_shen2021sar_net}.
Therefore, \textbf{multi-domain recommendation} (MDR) has recently become prevalent in industrial applications~\cite{m2m_zhang2022leaving_m2m}.
It aims to learn a unified model to perform predictions in multiple domains concurrently, utilizing the overlap of users and items among various domains~\cite{star_sheng2021one_star}.
To this end, it is significant for MDR models to model the commonalities and distinctions across different domains effectively~\cite{adasparse_yang2022adasparse}.
In recent years, numerous MDR models have been proposed and widely applied for multi-domain modeling using the hard-sharing~\cite{star_sheng2021one_star,hinet_zhou2023hinet,hamur_li2023hamur,hmoe_li2020improving,musenet_xu2023musenet} or soft-sharing~\cite{apg_yan2022apg,adasparse_yang2022adasparse,pepnet_chang2023pepnet,maria_tian2023multi,satrans_min2023scenario_satrans} paradigms.
Nevertheless, most existing methods only focus on the sophisticated design of model architectures.
Until now, there is still a lack of in-depth research on multi-domain problems from a more fundamental perspective of feature distributions.

\begin{figure}[!t]
  \vspace{-0.1in}
  \centering
    \includegraphics[width=0.9\linewidth]{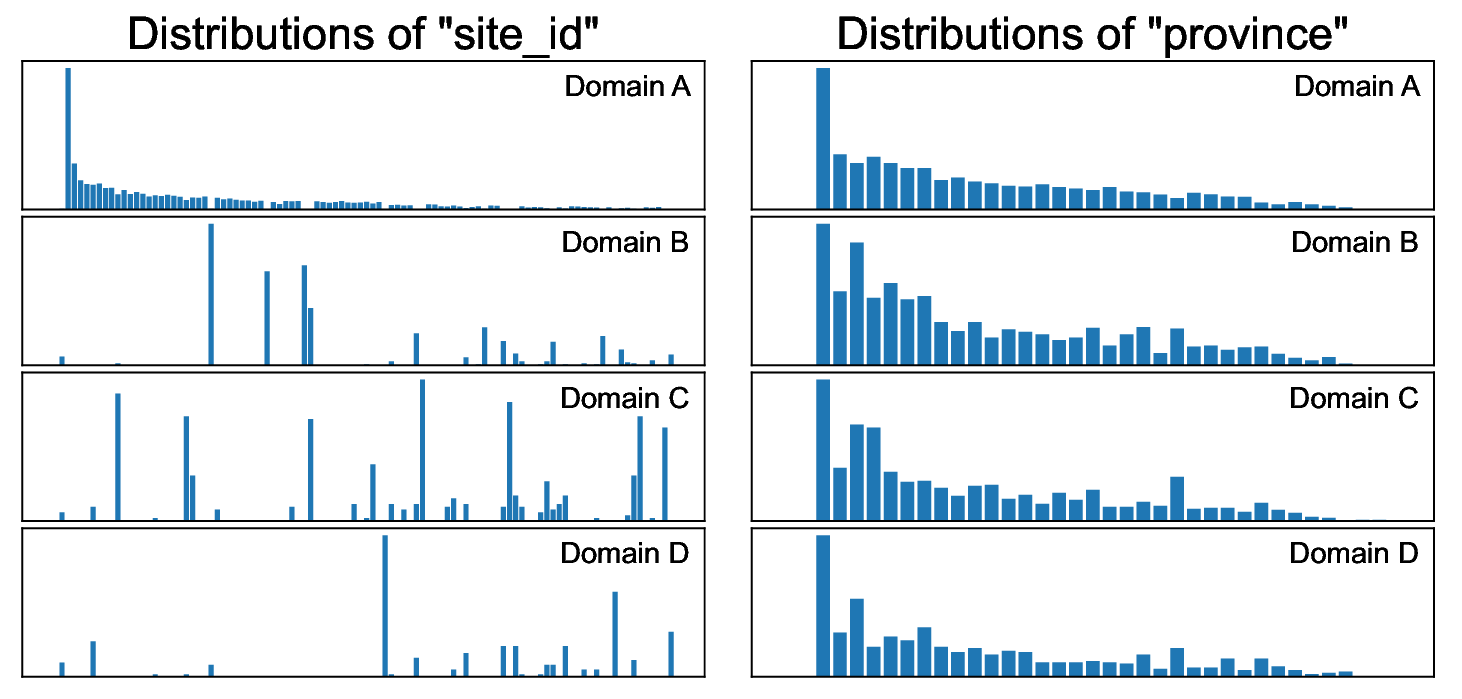}
  \vspace{-0.1in}
  \caption{Feature \textit{site\_id} and \textit{province} show diverse inter-domain differences in feature distributions.}
  \vspace{-0.2in}
  \label{fig.dist}
  \Description{}
\end{figure}

Generally, the differences across various domains primarily come from the distinct user populations and item collections.
From a technical perspective, the essence of multi-domain lies in the inter-domain differences in the distribution of features\footnote{In this paper, features refer to feature fields rather than values of a certain field.}.
The input of deep recommendation models typically contains numerous features~\cite{multisfs_wang2023single,autofield_wang2022autofield}.
If treating a particular feature as a random variable\footnote{For simplicity, we do not consider sequential features here. See Sec.~\ref{sec.dist_seq}.}, we can approximate its distribution using frequencies observed in the dataset.
Now, we examine the differences in feature distributions among different domains.
As shown in Fig.~\ref{fig.dist}, some features exhibit a similar distribution in each domain, like \textit{province}.
In contrast, significant inter-domain distribution differences exist in certain features, such as \textit{site\_id}.
We thereby suppose that domain distinctions are mainly caused by features like \textit{site\_id}, whose distributions are highly sensitive to domain changes.
We temporarily call them \textbf{domain-sensitive} features.
Conversely, features similar to \textit{province}, whose distributions remain almost the same across different domains, are considered to reflect the commonalities of domains.
Generally, the distribution change is slight when the domain varies for most features, so domain commonalities are relatively easy to learn for the model.
Only a few features are particularly sensitive to the domain.
Therefore, we hope that the MDR model can pay more attention to these domain-sensitive features and better capture domain distinctions.

\begin{figure}[!t]
  \centering
    \includegraphics[width=0.75\linewidth]{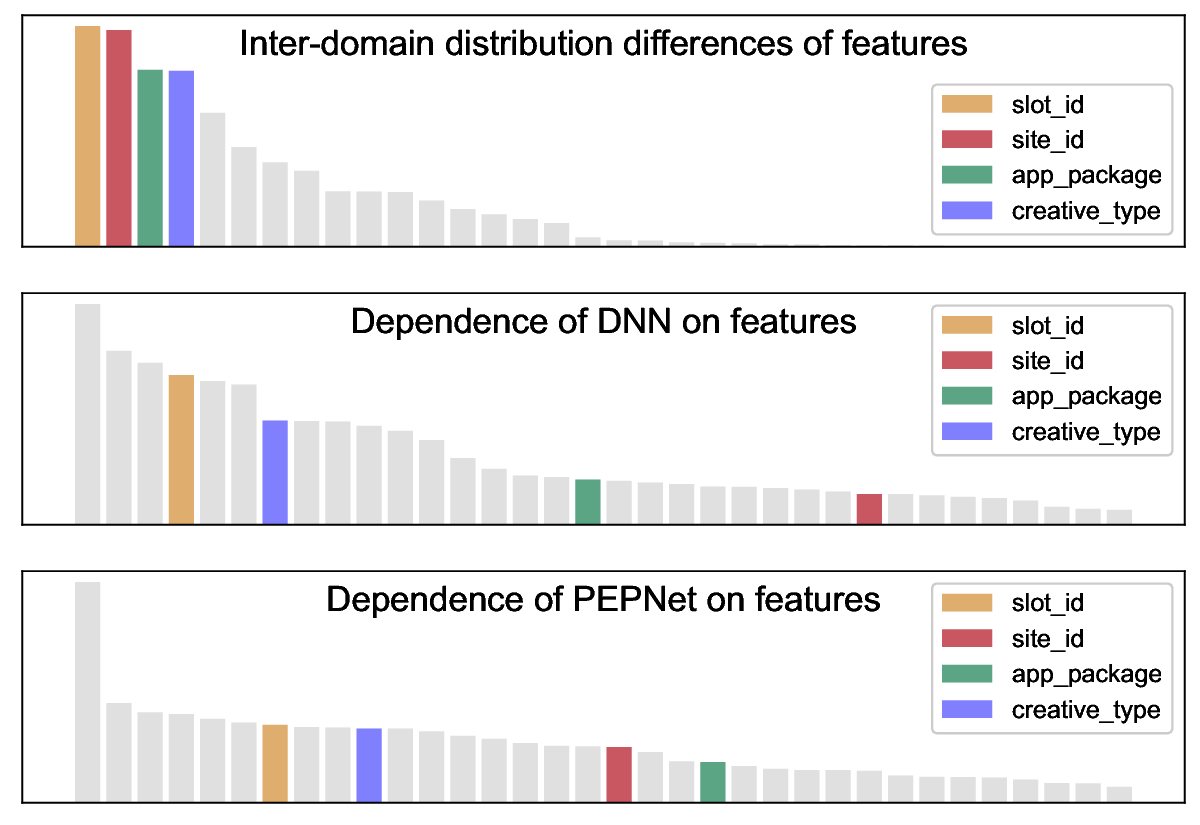}
  \vspace{-0.1in}
  \caption{Models neglect domain-sensitive features with the largest inter-domain distribution differences.}
  \vspace{-0.2in}
  \label{fig.neglect}
  \Description{}
\end{figure}

Existing MDR methods usually design specific model architectures for implicit multi-domain modeling~\cite{star_sheng2021one_star,hinet_zhou2023hinet,pepnet_chang2023pepnet,satrans_min2023scenario_satrans}.
We find through experiments that they may \textbf{neglect} domain-sensitive features, thus affecting the learning of domain distinctions.
Specifically, we separately train PEPNet~\cite{pepnet_chang2023pepnet}, a state-of-the-art MDR model, and DNN~\cite{dnn_covington2016deep} on our multi-domain industrial dataset.
Then we investigate their dependence on different features using Integrated Gradients~\cite{ig_sundararajan2017axiomatic_ig} (detailed in Sec.~\ref{sec.ig}), a popular feature attribution method for measuring the contribution of each feature to the model's predictions~\cite{attr_survey_guidotti2018survey,attr_survey_zhang2021survey}.
We sort all features according to the inter-domain distribution difference and the dependence of two models on them, respectively, to obtain three ordered feature lists, as shown in Fig.~\ref{fig.neglect}.
Then we select the top four features whose distributions are most sensitive to domains and highlight their positions in each list.
The four most domain-sensitive features rank lower in the list sorted by model dependence, indicating that models pay minor attention to these features.
This phenomenon implies that relying solely on specific model structures to capture domain distinctions may be unreliable.

In this paper, we propose a \textbf{Domain-Sensitive Feature Attribution} method to identify domain-sensitive features from the entire feature set.
The attribution method considers the inter-domain differences in both feature distributions and feature's effects on model predictions to ensure that the selected domain-sensitive features reflect domain-specific information.
Further, we design a concise and effective \textbf{Domain-Sensitive Feature Memory} architecture.
It takes domain-sensitive features as the input and extracts domain-specific information for the model to retrieve and utilize, thus preventing the neglect of domain-sensitive features and promoting learning domain distinctions.
We validate the effectiveness of our method through comprehensive offline and online experiments.

The main contributions of this paper are summarized as follows:
\begin{itemize}[leftmargin=*]
    \item We consider the multi-domain problem from the perspective of feature distributions and observe the neglect of domain-sensitive features in existing works.
    \item We propose a Domain-Sensitive Feature Attribution method. It identifies features that are sensitive to domain changes and can reflect domain distinctions based on feature distributions.
    \item We design a memory architecture to extract domain-specific information from domain-sensitive features for the model to retrieve and utilize. It emphasizes domain-sensitive features and improves multi-domain performance while maintaining reasonable computational efficiency. 
    \item We conduct extensive offline and online experiments to validate the effectiveness of our method. 
\end{itemize}

\section{Related Works}
\subsection{Multi-Domain Recommendation}
Multi-domain recommendation (MDR) aims to learn a unified model that simultaneously makes predictions for multiple domains, utilizing overlapping user and item information across different domains~\cite{star_sheng2021one_star}.
The recently proposed MDR methods can be divided into the following categories according to different multi-domain modeling paradigms.
(i) \textbf{Hard-sharing} methods~\cite{star_sheng2021one_star,hmoe_li2020improving,hinet_zhou2023hinet,hamur_li2023hamur,musenet_xu2023musenet,dadnn_he2020dadnn} usually employ a shared part at the bottom and structurally separate parts for each domain at the top of the model.
Multi-task learning methods, such as SharedBottom~\cite{sharedbottom_caruana1997multitask}, MMoE~\cite{mmoe_ma2018modeling_mmoe}, and PLE~\cite{ple_tang2020progressive_ple}, can be used for MDR by treating the prediction in each domain as an individual task.
STAR~\cite{star_sheng2021one_star} adds the Hadamard product interaction of model parameters between the shared and independent parts of SharedBottom.
HiNet~\cite{hinet_zhou2023hinet} adopts the variants of MMoE with attention-based gating mechanisms.
(ii) \textbf{Soft-sharing} methods~\cite{apg_yan2022apg,adasparse_yang2022adasparse,pepnet_chang2023pepnet,satrans_min2023scenario_satrans,maria_tian2023multi,m2m_zhang2022leaving_m2m} utilize a globally shared network and dynamically adjust its architecture, parameters, or hidden representations, conditioned on domains or samples.
AdaSparse~\cite{adasparse_yang2022adasparse} adjusts the model architecture by pruning neurons.
APG~\cite{apg_yan2022apg} and M2M~\cite{m2m_zhang2022leaving_m2m} adaptively generate model parameters based on meta-learning~\cite{meta_hospedales2021meta}. 
PEPNet~\cite{pepnet_chang2023pepnet} adopts gating networks to adjust the distribution of feature embeddings and the model's hidden representations, inspired by LHUC ~\cite{lhuc_swietojanski2016learning}.
In summary, existing works have designed elaborate model architectures for MDR. 
However, none of them has considered the multi-domain problem from a more fundamental perspective of feature distributions.

\subsection{Feature Attribution for Neural Networks}
Feature attribution refers to quantitatively attributing the predictions of a neural network to its input features~\cite{attr_survey_guidotti2018survey, attr_survey_zhang2021survey, 14_attr_deng2024unifying_14}.
Through feature attribution, we can measure a feature's contribution to the model's output, i.e., its importance~\cite{ig_sundararajan2017axiomatic_ig}.
The commonly used feature attribution methods can be divided into the following categories:
(i) \textbf{Gradient-based} methods, e.g., SynFlow~\cite{synflow_tanaka2020pruning_synflow}, Integrated Gradients~\cite{ig_sundararajan2017axiomatic_ig}, and Grad-CAM~\cite{grad_cam_selvaraju2017grad_cam}, calculate the gradient of the model's output w.r.t. the input features to determine their importance.
(ii) \textbf{Perturbation-based} methods, e.g., Occlusion-1~\cite{occlusion_1_zeiler2014visualizing_occlusion_1} and Occlusion-patch~\cite{occlusion_patch_zintgraf2017visualizing_occlusion_patch}, perturb or modify specific features and observe the changes in the model's output to infer the importance of each feature.
(iii) \textbf{Back-propagation} methods, e.g., LRP~\cite{lrp_bach2015pixel_lrp}, DeepLIFT~\cite{deeplift_shrikumar2017learning_deeplift}, and Deep SHAP~\cite{deepshap_lundberg2017unified_deepshap}, use a technique that estimates the attribution of intermediate variables in the output layer and propagates it back to the input layer by layer.
Among the methods above, the gradient-based Integrated Gradients~\cite{ig_sundararajan2017axiomatic_ig} is theoretically intuitive and has a concise implementation, making it easy to integrate into large-scale recommendation systems.

\section{preliminaries}
This paper focuses on the multi-domain advertisement ranking problem.
Assuming there are $K$ domains, the data from each domain shares the same feature space $\mathcal{X}$ and label space $\mathcal{Y}$ but with different feature distributions. 
The feature set includes user features, item features, and context features. 
There is also a domain feature indicating which domain the sample belongs to.
Labels are binary and represent the state of user feedback, such as clicks and conversions.
Donate the dataset of the $k$-th domain as $\mathcal{D}^k=\{(x_i^k,y_i^k)\}_{i=1}^{|\mathcal{D}^k|},k=1,2,\dots,K$,
where $x_i^k\in\mathcal{X}$ and $y_i^k\in\mathcal{Y}$ are the features and label of the $i$-th sample.
Multi-domain ad ranking aims to construct and train a unified predicting model $\hat{y}=F(x;\Theta)$ with parameter set $\Theta$ using $\mathcal{D}=\bigcup_{k=1}^K{\mathcal{D}^k}$ and expects $F(\cdot)$ to make accurate predictions in the multi-domain test dataset $\mathcal{D}_\text{test}$.
Using binary cross-entropy loss $l(\cdot,\cdot)$, the optimization objective is
\begin{equation}
    \mathcal{L}=\frac{1}{|\mathcal{D}|}\sum_{k=1}^{K}{\sum_{i=1}^{|\mathcal{D}^k|}{l(y_i^k,\hat{y}_i^k)}}.
\end{equation}

\section{methodology}
Our method mainly consists of two parts. 
First, we propose an attribution method to identify domain-sensitive features based on feature distributions and their effects on model predictions. 
Second, we design a memory architecture to extract domain-specific information from domain-sensitive features and enhance the model's awareness of domain distinctions, thus improving multi-domain predicting performances.

\subsection{Domain-Sensitive Feature Attribution}
\subsubsection{Introduction of Domain-Sensitive Features}
Firstly, we define domain-sensitive features as features that exhibit significant differences across various domains.
The differences emerge from two aspects: the \textbf{feature distribution} and the \textbf{feature's effect} on model predictions.
There are overlaps and differences among user groups, item collections, and domain contexts within different domains, reflecting domain commonalities and distinctions.
After abstracting the information of users, items, and domain contexts as features, differences in feature distributions can reflect domain distinctions intuitively.
On the other hand, it is also necessary to consider the feature's effect on model predictions because analyzing feature distributions alone can not establish the relationship between feature and model predictions. 
Consider the following two extreme cases:
\begin{itemize}[leftmargin=*]
\item 
A certain feature has a large inter-domain distribution difference, while the model disregards it. 
The feature is then mistakenly believed to be domain-sensitive, although it cannot deliver domain distinctions to the model.
\item 
A certain feature has exactly the same distribution in each domain, but the effect of its different values on model predictions varies significantly. 
Intuitively, this feature should have been domain-sensitive, but we consider it completely insensitive.
\end{itemize}
Consequently, we also consider the feature's effect on model predictions.
Overall, our attribution method aims to identify domain-sensitive features that demonstrate significant inter-domain differences in both distributions and effects on model predictions.

\subsubsection{Effect-Weighted Feature Distribution}

Assume there are $m$ features $f_1,f_2,\dots,f_m$ in the feature set.
For the $i$-th sample $(x_i^k,y_i^k)\in \mathcal{D}^k$ belonging to the $k$-th domain, the input $x_i^k=(x_{i,1}^k,x_{i,2}^k,\dots,x_{i,m}^k)$ consists of $m$ values, each corresponding to a feature. 
We consider a certain categorical\footnote{Numerical features need to be discretized first.} feature $f_j$ with value set $V_j=\{v_{j,1},v_{j,2},\dots,v_{j,{n_j}}\}$, where $n_j$ is the number of possible values. 
Regarding $f_j$ as a discrete random variable, the distribution of $f_j$ in domain $k$ can be estimated from the dataset, that is
\begin{equation}
\label{eq.dist_func}
    P^k_{f_j}(v_{j,r})=\frac{1}{|\mathcal{D}^k|}\sum_{i=1}^{|\mathcal{D}^k|}{\mathbb{I}(x^k_{i,j}=v_{j,r})},r=1,2,\dots,n_j,
\end{equation}
where $\mathbb{I}(\cdot)$ is the 0-1 indicator function.

Next, we consider the effect of features on model outputs.
Generally, the effect of a certain feature's value in different samples varies. 
Based on the feature attribution theory of neural networks~\cite{attr_survey_guidotti2018survey, attr_survey_zhang2021survey}, we can quantitatively measure a feature's effect in each sample and obtain a corresponding attribution score.
We introduce the attribution scores to Eq.~(\ref{eq.dist_func})  and define the \textbf{effect-weighted feature distribution} of $f_j$ as
\begin{equation}
    \tilde{P}^k_{f_j}(v_{j,r})=\frac{1}{Z_j^k}\sum_{i=1}^{|\mathcal{D}^k|}{a_{i,j}^k\cdot\mathbb{I}(x^k_{i,j}=v_{j,r})},r=1,2,\dots,n_j,
    \label{eq.attribution_distribution}
\end{equation}
where $a_{i,j}^k$ is the attribution score of feature $f_j$ with value $v_{j,r}$ in input $x_i^k$, and $Z_j^k$ is a normalization constant satisfying $Z_j^k=\sum_{i=1}^{|\mathcal{D}^k|}{a_{i,j}^k}$.
Eq.~(\ref{eq.attribution_distribution}) can be understood as an effect distribution, i.e., the distribution of effects on model outputs of feature $f_j$'s possible values in domain $k$.
It both reflects the feature distribution and considers the effect of different feature values on the model output.

\subsubsection{Feature Attribution for Measuring Effects}\label{sec.ig}
We adopt Integrated Gradients~\cite{ig_sundararajan2017axiomatic_ig} (IG) to attributing features and obtain attribution scores.
Given a trained model $F(\cdot)$, an input $x$, and a baseline input $x^\prime$ representing information absence, IG is defined as the path integral of the gradients along the straight line from $x^\prime$ to $x$,
that is,
\begin{equation}
    \text{IG}_j(x,x^\prime)=(x_j-x^\prime_j)\cdot\frac{1}{T}\sum_{t=1}^{T}{\frac{\partial F(x^\prime+\frac{t}{T}(x-x^\prime))}{\partial x_j}}
    \label{eq.ig}
\end{equation}
where $x_j,x^\prime_j$ are the values of the $j$-th feature $f_j$, $T$ is the number of interpolation steps.
In deep recommendation models, input features are typically transformed into embeddings~\cite{dnn_covington2016deep}.
In this case, $\text{IG}_j(x,x^\prime)$ is a vector, so we sum it up to get a scalar attribution score of $f_j$. 
To determine the value of attribution score $a_{i,j}^k$ in Eq.~(\ref{eq.attribution_distribution}), we first train a DNN model on the training dataset, and then utilize it to calculate $a_{i,j}^k$ through Eq.~(\ref{eq.ig}), i.e.,
\begin{equation}
    \label{eq.attribution_with_ig}
    a_{i,j}^k=\text{IG}_j(x_i^k, \mathbf{0}),
\end{equation}
where $\mathbf{0}$ is a zero vector\footnote{For embedding inputs, the baseline can be zero vectors, following the original paper.} with the same shape as the input $x_i^k$.

\subsubsection{Generalized Distribution of Sequential Features}\label{sec.dist_seq}
So far, we have only considered scalar features, while sequential features are also frequently used in recommendations~\cite{din_zhou2018deep,dien_zhou2019deep}.
However, the variable length of sequential features prevents the direct definition of their distributions.
Here, we introduce a generalization such that each sequential feature can obtain a distribution similar to Eq.~(\ref{eq.attribution_distribution}).
Consider a certain sequential feature $f_{j^\prime}$ with the value set $V_{j^\prime}=\{v_{j^\prime,1},v_{j^\prime,2},\dots,v_{j^\prime,n_{j^\prime}}\}$.
Given the $i$-th sample in the $k$-th domain $x_i^k$, the value of 
sequential feature $f_{j^\prime}$ is a sequence $x^k_{i,j^\prime}=(x^k_{i,j^\prime,1},x^k_{i,j^\prime,2},\dots,x^k_{i,j^\prime,L^k_{i,j^\prime}})$,
where $L^k_{i,j^\prime}$ is the sequence length.
Denote $a^k_{i,j^\prime}$ as the attribution score of $x^k_{i,j^\prime}$, we define the effect-weighted distribution of sequential feature $f_{j^\prime}$ in domain $k$ to be
\begin{equation}\small
\begin{aligned}
    \tilde{P}^k_{f_j^\prime}(v_{j^\prime,r})=\frac{1}{Z^k_{j^\prime}}\sum_{i=1}^{|\mathcal{D}^k|}{\Big[a^k_{i,j^\prime}\cdot\frac{1}{L^k_{i,j^\prime}}\sum_{l=1}^{L^k_{i,j^\prime}}{\mathbb{I}(x^k_{i,j^\prime,l}=v_{j^\prime,r})}\Big]},
    r=1,2,\dots,n_{j^\prime},
\end{aligned}
    \label{eq.attr_dist_seq}
\end{equation}
where $Z_{j^\prime}^k$ is a normalization constant and $1/L^k_{i,j^\prime}$ is a factor to eliminate the effect of different sequence length.
Eq.~(\ref{eq.attr_dist_seq}) can be interpreted as the process of accumulating each occurrence of feature value $v_{j^\prime,r}$ into the distribution with a weight of $a^k_{i,j^\prime}$.

\begin{algorithm}[!t]
    \caption{Domain-sensitive feature attribution}
    \label{alg.attribution}
    \algsetup{linenosize=\normalsize}
    \begin{algorithmic}[1]
        \REQUIRE multi-domain training dataset $\mathcal{D}$, \\ \hspace{1.3em} feature set $\mathcal{F}=\{f_1,f_2,\dots,f_m\}$, number of domains $K$
        \ENSURE domain-sensitivity of each feature

        \STATE Train a base DNN model $\hat{y}=F(x)$ on $\mathcal{D}$
        \STATE Approximate feature attribution scores of each sample in $\mathcal{D}$ using the trained model $F(\cdot)$ via Eq.~(\ref{eq.attribution_with_ig})
        \FOR{each feature $f_j\in\mathcal{F}$}
            \FOR{domain $k=1,2,\dots,K$}
                \STATE Calculate the effect-weighted distribution $\tilde{P}_{f_j}^k$ via Eq.~(\ref{eq.attribution_distribution}) or Eq.~(\ref{eq.attr_dist_seq}) based on whether $f_j$ is scalar or sequential.
            \ENDFOR
            \STATE Calculate the domain-sensitivity of $f_j$ via Eq.~(\ref{eq.domain_sensitivity}) using JS divergence in Eq.~(\ref{eq.js_divergence}) or Wasserstein distance in Eq.~(\ref{eq.w_dis}) based on whether $f_j$ is categorical or numerical.
        \ENDFOR
    \end{algorithmic}
\end{algorithm}

\subsubsection{Measurement of Distribution Distances}

Currently, we have obtained the distribution of each feature in different domains.
In this section, we measure the inter-domain distribution differences of each feature.
The case of categorical and numerical features will be discussed separately.
We use the Jensen–Shannon (JS) divergence for categorical features, i.e.,
\begin{equation}
    \label{eq.js_divergence}
    \text{JS}(P,Q)=\frac{1}{2}\sum_x{P(x)\log\frac{P(x)}{M(x)}}+\frac{1}{2}\sum_x{Q(x)\log\frac{Q(x)}{M(x)}},
\end{equation}
where $P$ and $Q$ are two discrete possibility distributions, and $M=(P+Q)/2$ is a mixture distribution.
For numerical features, the JS divergence, which measures the overlap of distributions, cannot capture the quantitative relationship among different values.
For instance, consider the following three degenerate distributions $\lambda,\mu,\nu$ without overlaps: $P_\lambda(x=0)=1,P_\mu(x=0.1)=1,P_\nu(x=1)=1$.
Intuitively, the distance between $\mu$ and $\lambda$ should be smaller than between $\nu$ and $\lambda$, as 0.1 is closer to 0 than 1. 
Nevertheless, the JS divergence between each pair of them is identical.
Inspired by WGAN~\cite{wgan_arjovsky2017wasserstein_wgan}, we use the Wasserstein distance~\cite{wdis_kantorovich1960mathematical_wdis} to measure the distribution distance of numerical features.
For two distributions $P$ and $Q$, the Wasserstein distance\footnote{Refer to the most commonly used 1-order Wasserstein distance.} is defined by the formula
\begin{equation}
W(P,Q)=\mathop{\text{inf}}\limits_{\gamma\in\Pi[P,Q]}\iint{\gamma(x,y)d(x,y)\;\text{d}x\text{d}y},
    \label{eq.w_dis}
\end{equation}
where $d(\cdot,\cdot)$ is a metric function, and $\Pi[P,Q]$ denotes all joint distributions that have margins $P$ and $Q$.
We use $L^1$-norm as the metric function. 
The Wasserstein distance can be succinctly described as the minimum cost required to transport one distribution to another.
In particular, if $P$ and $Q$ are two one-dimensional empirical distributions, Eq.~(\ref{eq.w_dis}) can be solved in $O(n\log n)$.
For convenience, we use the POT~\cite{pot_flamary2021pot} package to solve it.
For the above degenerate distributions $\lambda,\mu,\nu$, we have $W(\mu,\lambda)=0.1$ and $W(\nu,\lambda)=1$, which reflects the geometric distance relationship beyond overlaps.

Given a certain feature $f_j,j\in\{1,2,\dots,m\}$ and its effect-weighted distributions in $K$ domains $\tilde{P}_{f_j}^1, \tilde{P}_{f_j}^2,\dots,\tilde{P}_{f_j}^K$, we define the \textbf{domain-sensitivity} (DS) of feature $f_j$ to be
\begin{equation}
    \text{DS}_j=\sum_{k=1}^{K}{\sum_{k^\prime=k+1}^{K}{\text{Dist}(\tilde{P}_{f_j}^k,\tilde{P}_{f_j}^{k^\prime})}},
    \label{eq.domain_sensitivity}
\end{equation}
where $\text{Dist}(\cdot,\cdot)$ denotes the distribution distance function.
$\text{Dist}(\cdot,\cdot)$ is the JS divergence for categorical features, while it is the Wasserstein distance for numerical features.
Domain-sensitive feature attribution is essentially calculating DS of each feature, and its complete procedure is described in Algorithm~\ref{alg.attribution}.  

Features with relatively large domain-sensitivity are considered to carry more domain distinctions.
Therefore, we rank features according to their domain-sensitivity and select the top several features as domain-sensitive features.
Based on whether they are categorical or numerical and scalar or sequential, we group the features and rank each group separately to avoid the influence of mismatched numerical range.

\subsection{Domain-Sensitive Feature Memory}\label{sec:memory}
\begin{figure}[!t]
  \centering
    \includegraphics[width=0.99\linewidth]{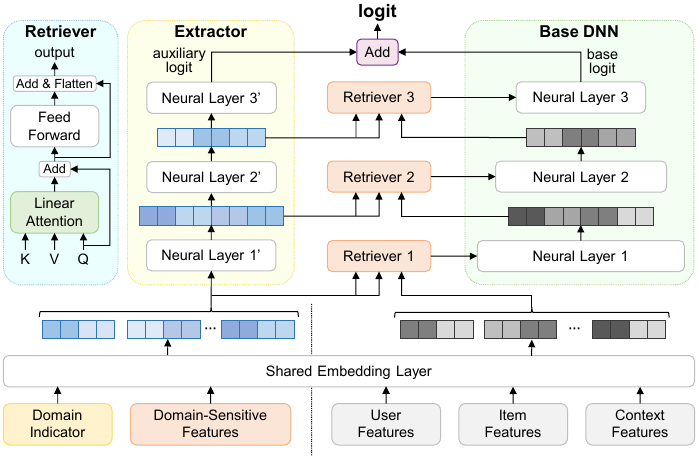}
  \vspace{-0.1in}
  \caption{The architecture of Domain-Sensitive Feature Memory consists of two key modules: the Extractor and Retriever.}
  \vspace{-0.15in}
  \label{fig.model}
  \Description{}
\end{figure}

In this section, we propose a concise yet effective model architecture, i.e., the retrievable Domain-Sensitive Feature Memory, to emphasize domain-sensitive features and improve multi-domain predicting performance. 
It extracts domain-specific information from domain-sensitive features for the base model to retrieve and utilize, enhancing the awareness of domain distinctions. 
As illustrated in Fig.~\ref{fig.model}, the overall structure is a dual-tower with a shared embedding layer.
One tower is a base model, and another is a Extractor, with Retrievers connecting the two towers. 
For simplicity, we use DNN~\cite{dnn_covington2016deep} as the base model here, which can be replaced by a general model.

\subsubsection{Domain-Sensitive Feature Extractor}

The Extractor is a DNN-based auxiliary network parallel to the base network and takes domain-sensitive features as input.
It progressively extracts information from domain-sensitive features.
Let $e_1,e_2,\dots,e_m$ denote the embeddings of features $f_1,f_2,\dots,f_m$.
Assume the selected domain-sensitive features are $f_{s_1},f_{s_2},\dots,f_{s_{n_s}}$, the Extractor's input is $E_\text{ext}=e_{s_1}\oplus e_{s_2}\oplus \dots\oplus e_{s_{n_s}}$,
where $\oplus$ denotes a concatenating operation.
The output logit is 
\begin{align}
    \label{eq.extractor_output}
    \begin{aligned}
    &l_\text{ext}=\text{DNN}_\text{ext}(E_\text{ext}).
    \end{aligned}
\end{align} 
We merge the logit of the two towers as the final output to directly emphasize domain-sensitive features, i.e., $l=l_\text{base}+l_\text{ext}$,
where $l_\text{base}$ is the logit of the base model.

\subsubsection{Domain-Specific Information Retriever}

In order to emphasize domain-sensitive features and maintain awareness of domain distinctions,
we propose a Domain-Specific Information Retriever utilizing the linear cross-attention~\cite{linear_attn_katharopoulos2020transformers_linear_attention}. 
Assume both of the two towers have $L$ neural layers.
Between every two corresponding layers, we incorporate a Retriever to retrieve the domain-specific information extracted from domain-sensitive features and integrate it into the base model with dynamic weights. 
For the embedding layer, the original embeddings feed to the base model and the Extractor is $\{e_{1}, e_{2}, \dots, e_{m}\}$ and $\{e_{s_1}, e_{s_2}, \dots, e_{s_{n_s}}\}$, respectively.
We pack them into two matrices $Z\in\mathbb{R}^{m\times d}$ and $Z_\text{ext}\in\mathbb{R}^{n_s\times d}$, where $d$ is the embedding size.
Then the cross-attention output is
\begin{equation}
    A=\text{Attention}(Q,K,V)W_O,
\end{equation}
\begin{equation}
Q=ZW_Q,K=Z_\text{ext}W_K,V=Z_\text{ext}W_V,
\end{equation}
where $W_Q,W_K,W_V\in\mathbb{R}^{d\times d^\prime}$ and $W_O\in\mathbb{R}^{d^\prime\times d}$ are parameter matrices for projections, and $d^\prime$ is the hidden size of attention. 
The attention output $A$ carries domain distinctions retrieved from domain-sensitive features,
while the original input $Z$ contains domain commonalities.
We add them up and apply a two-layer fully connected feed-forward network with a residual connection~\cite{residual_he2016deep_residual}, i.e.,
\begin{equation}
Z^A=Z+A,\ \ Z^\prime=\text{FFN}(Z^A)+Z^A,
\end{equation}
\begin{equation}
\text{FFN}(x)=\text{ReLU}(xW_1+b_1)W_2+b_2.
\end{equation}
Then $Z^\prime$ is feed to the base model.

In addition, we also employ the Retriever between each hidden layer of the dual-tower to retrieve refined domain-specific information from the Extractor's hidden representations, ensuring the base model's continual awareness of domain distinctions. 
While the hidden representations of DNNs are one-dimensional vectors, treating them as sequences with a token dimension of 1 enables direct utilization of attention modules. 

\subsubsection{Linear Attention for Low Complexity}
One major limitation of the vanilla softmax attention~\cite{attention_vaswani2017attention} is its quadratic computational complexity with respect to the sequence length~\cite{linear_attn_katharopoulos2020transformers_linear_attention,efficien_attention_shen2021efficien_attention,linformer_wang2020linformer,performer_choromanski2020rethinking_performer}. 
The hidden size of DNNs is typically in the order of hundreds.
When applying softmax attentions to the hidden representation of DNNs, the quadratic complexity could lead to a slow computational speed and harm scalability.
Therefore, we adopt the linear attention~\cite{linear_attn_katharopoulos2020transformers_linear_attention} as a substitute,
which is defined by
\begin{equation}
\label{eq.linear_attention}
    \text{LinearAttention}(Q,K,V)_i=\frac{\sum_{j=1}^{n}{\phi(Q_i)\phi(K_j)^\text{T}V_j}}{\sum_{j=1}^{n}{\phi(Q_i)\phi(K_j)^\text{T}}},
\end{equation}
where subscript $i$ indicates the $i$-th row of the matrix, $\phi(x)=\text{ELU}(x)+1$ is the feature map, and $\text{ELU}(\cdot)$ denotes the Exponential Linear Unit function~\cite{elu_clevert2015fast_elu}.
Let $n$ be the sequence length and $d$ be the token size, Eq.~(\ref{eq.linear_attention}) allows a complexity of $O(nd^2)$ while the complexity of softmax attention is $O(n^2d)$.
We have $d \ll n$ here, so the linear attention can significantly improve computational efficiency, thus allowing online deployment.

\section{Experiments}

\begin{table*}[!t]
\centering
\vspace{-0.1in}
\small
\caption{The overall performance of each method in terms of AUC.}
\vspace{-0.1in}
\label{tab:main_auc}
\begin{tabular}{c|c|cc|cccc|ccc|c}
    \toprule
    \multirow{3}{*}{\textbf{Dataset}} & \multirow{3}{*}{\textbf{Domain}} & \multicolumn{10}{c}{\textbf{Methods}} \\ 
    \cline{3-12}
    ~ & ~ & \multicolumn{2}{c|}{Naive} & \multicolumn{4}{c|}{Hard-sharing} & \multicolumn{3}{c|}{Soft-sharing} & \multirow{3}{*}{Ours} \\ 
    \cline{3-11}
    ~ & ~ & \makecell[c]{Shared\\DNN} & \makecell[c]{Separate\\DNN} & \makecell[c]{Shared\\Bottom} & MMoE & STAR & HiNet & APG & AdaSparse & PEPNet \\ 
    \hline
    \multirow{5}{*}{\textbf{Industrial}} 
      & A       & 0.8245 & 0.8268 & 0.8261 & 0.8277 & 0.8286 & 0.8269 & 0.8284 & 0.8259 & \underline{0.8293} & \textbf{0.8309} \\ 
    ~ & B       & 0.8307 & 0.8313 & 0.8329 & \underline{0.8336} & 0.8335 & \underline{0.8336} & 0.8320 & 0.8313 & 0.8335 & \textbf{0.8340} \\ 
    ~ & C       & 0.8184 & 0.8255 & 0.8272 & \underline{0.8328} & 0.8323 & 0.8308 & 0.8262 & 0.8205 & 0.8253 & \textbf{0.8331} \\ 
    ~ & D       & 0.8774 & 0.8799 & 0.8818 & \underline{0.8852} & 0.8828 & 0.8828 & 0.8819 & 0.8812 & 0.8803 & \textbf{0.8862} \\ 
    ~ & Overall & 0.8423 & 0.8440 & 0.8445 & 0.8461 & 0.8464 & 0.8452 & 0.8457 & 0.8437 & \underline{0.8466} & \textbf{0.8484} \\ 
    \hline
    \multirow{4}{*}{\textbf{Ali-CCP}} 
      & E       & 0.6208 & 0.6210 & 0.6224 & 0.6223 & 0.6215 & 0.6231 & 0.6222 & 0.6228 & \underline{0.6241} & \textbf{0.6264} \\ 
    ~ & F       & 0.6228 & 0.6211 & 0.6250 & 0.6254 & 0.6250 & 0.6260 & 0.6252 & 0.6258 & \underline{0.6267} & \textbf{0.6281} \\ 
    ~ & G       & 0.5996 & 0.5683 & 0.6012 & 0.6024 & 0.5972 & \underline{0.6031} & 0.6002 & 0.6025 & 0.6019 & \textbf{0.6037} \\ 
    ~ & Overall & 0.6216 & 0.6209 & 0.6229 & 0.6236 & 0.6227 & 0.6244 & 0.6234 & 0.6241 & \underline{0.6252} & \textbf{0.6272} \\ 
    \bottomrule
\end{tabular}
\vspace{-0.1in}
\end{table*}

In this section, we conduct extensive experiments on different datasets and the online A/B test to validate the effectiveness of our method and answer the following research questions.

\begin{itemize}[leftmargin=*]
    \item RQ1: How does our method perform compared with state-of-the-art methods?
    \item RQ2: Does the proposed memory architecture effectively enhance domain-sensitive features? How does each part contribute to the performance? 
    \item RQ3: Do domain-sensitive features obtained by the attribution method effectively promote the learning of domain distinctions and improve multi-domain recommendation performances?
    \item RQ4: How does our method compare with baselines in terms of computational efficiency?
    \item RQ5: How does the proposed method perform in the real-world recommender system?
\end{itemize}

\subsection{Experiment Setup}
\subsubsection{Datasets and Metrics}
We conduct comparative experiments on public and industrial datasets. 
The public dataset is Ali-CCP\footnote{\href{https://tianchi.aliyun.com/dataset/408}{https://tianchi.aliyun.com/dataset/408}}, a widely used multi-task e-commerce recommendation dataset with three domains.
We choose \textit{click} as the target and exclude sequential features here.
The industrial dataset comes from Huawei's online advertising platform, and the target is \textit{conversion}.
We divide it into four domains based on ad types.
The detailed statistics of datasets are shown in Table~\ref{tab:data_stats}.
We evaluate the performance of models using the most commonly used AUC, the area under the ROC curve.  
In order to better demonstrate the performance in multiple domains, we also provide the AUC within each domain separately. 
Each reported result is the average result of five randomized experiments.

\begin{table}[!h]
  \small
  \centering
  \vspace{-0.05in}
  \caption{Data statistics by domain.}
  \vspace{-0.1in}
  \label{tab:data_stats}
  \resizebox{\linewidth}{!}{
  \begin{tabular}{c|c|cccccc}
    \toprule
    \textbf{Dataset} & \textbf{Domain} & \textbf{\#Sample} & \textbf{Prop.} & \textbf{Pos. Rate} & \textbf{\#Train} & \textbf{\#Valid} & \textbf{\#Test} \\
    \midrule
    \multirow{5}{*}{\makecell[c]{\textbf{Industrial}\\(115 features)}} 
    & A         & 30.4M & 80.62\% & 1.10\% & 26.2M & 2.36M & 1.87M \\
    & B         & 3.47M & 9.20\%  & 3.61\% & 3.03M & 0.26M & 0.18M \\
    & C         & 2.38M & 6.30\%  & 0.98\% & 2.08M & 0.18M & 0.12M \\
    & D         & 1.46M & 3.88\%  & 2.32\% & 1.29M & 0.11M & 0.06M \\
    & Overall   & 37.7M & -       & 1.37\% & 32.6M & 2.91M & 2.23M \\
    \midrule
    \multirow{4}{*}{\makecell[c]{\textbf{Ali-CCP}\\(19 features)}}    
    & E         & 52.4M & 61.46\% & 3.81\% & 23.5M & 2.61M & 26.3M \\
    & F         & 32.2M & 37.79\% & 4.00\% & 14.3M & 1.59M & 16.4M \\
    & G         & 0.64M & 0.75\%  & 4.38\% & 0.29M & 0.03M & 0.32M \\
    & Overall   & 85.3M & -       & 3.89\% & 38.1M & 4.23M & 43.0M \\
  \bottomrule
\end{tabular}
}
\vspace{-0.15in}
\end{table}

\subsubsection{Baselines}
We compare the performances of different mainstream baselines, which can be divided into: (i) \textbf{Naive} methods: Separate DNN, i.e., training DNNs separately for different domains; Shared DNN, i.e., training a unified DNN using samples from all domains; (ii) \textbf{Hard-sharing} methods: SharedBottom~\cite{sharedbottom_caruana1997multitask}, MMoE~\cite{mmoe_ma2018modeling_mmoe}, STAR~\cite{star_sheng2021one_star}, and HiNet~\cite{hinet_zhou2023hinet}; (iii) \textbf{Soft-sharing} methods: APG~\cite{apg_yan2022apg}, AdaSparse~\cite{adasparse_yang2022adasparse}, and PEPNet~\cite{pepnet_chang2023pepnet}. 
Among them, HiNet and PEPNet are the state-of-the-art multi-domain recommendation models.

\subsubsection{Hyperparameter Setup}
For the industrial dataset, the model's hidden size is [256, 128, 64], and the batch size is 20480.
The hidden size of cross-attentions is set to 32 and 8 for the embedding and hidden layers. 
We use the top-5 domain-sensitive categorical features.
For the public dataset, the model's hidden size is [128, 64, 32], and the batch size is 4096. 
The hidden size of cross-attentions is 16 for the embedding layer and 4 for the hidden layers.
We use the top-1 domain-sensitive feature \textit{user\_id}.
For both datasets, the embedding size is 8, the optimizer is Adam with a learning rate of $10^{-3}$, and the number of interpolation steps $T$ in Eq.~(\ref{eq.ig}) is set to 5.

\subsection{Overall Performance (RQ1)}

Table~\ref{tab:main_auc} presents the performance of our method and baselines in terms of AUC on industrial and public datasets.
Our method outperforms all baselines in both overall and domain-wise performance on the two datasets. 
Limited by space, we mainly analyze the results on our industrial dataset, while the results on Ali-CCP are similar.
Domain A is a dominant domain which constitutes over 80\% of the entire dataset.
The performance on Domain A reflects how well the model has learned the domain commonalities.
Domains B, C, and D are tail domains with a small portion of samples, requiring the model to capture domain distinctions well.
A comparison with two naive methods reveals that a completely shared model may get confused about domains and result in poor performance across all domains.
Hard-sharing models with individual domain-specific components allow more effective learning of domain distinctions, hence performing better in three tail domains than soft-sharing models.
On the contrary, soft-sharing models using an entirely shared architecture can help learn domain commonalities but may overlook domain distinctions.
Therefore, soft-sharing models perform better in the dominant domain A than hard-sharing models but worse in the tail domains.
Our method can also be categorized into a soft-sharing model.
By introducing a domain-sensitive feature memory to extract and enhance domain-specific information, our method adequately learns both domain commonalities and distinctions, resulting in superior performance across all domains.

\subsection{Ablation Study (RQ2)}
\begin{figure}[!h]
  \centering
    \includegraphics[width=0.99\linewidth]{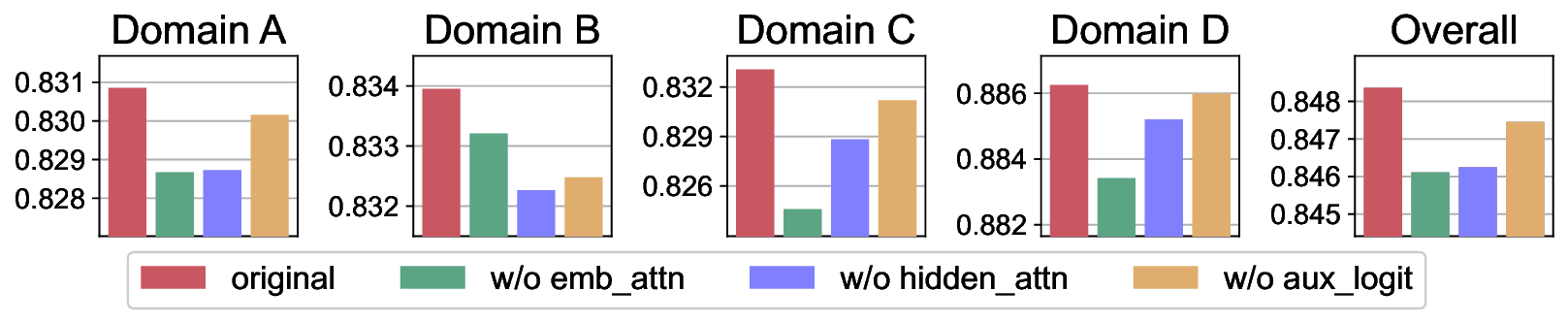}
  \vspace{-0.1in}
  \caption{Results (AUC) of the ablation study on several variants of our domain-sensitive feature memory.}
  \vspace{-0.2in}
  \label{fig.ablation_module}
  \Description{}
\end{figure}

In this section, we conduct an ablation study on the industrial dataset to validate the effectiveness of each component in the proposed domain-sensitive feature memory architecture.
We remove the Retriever after the embedding layer (w/o emb\_attn), Retrievers between the hidden layers (w/o hidden\_attn), and the auxiliary logit summation (w/o aux\_logit) from the model, respectively.
As shown in Fig.~\ref{fig.ablation_module}, removing Retrievers in either embedding or hidden layers can decrease overall and domain-wise AUC, especially for tail domains like C and D, 
This indicates that the Extractor can effectively extract domain-specific information, and the Retriever can properly retrieve it for the model to utilize.
Furthermore, removing the auxiliary logit also results in an AUC decrease, suggesting that the logit output by the Extractor can also assist the model in capturing domain distinctions.

\subsection{\hspace{-0.85em}Emphasizing Domain-Sensitive Features (RQ2)}
\begin{figure}[!h]
  \centering
    \includegraphics[width=0.9\linewidth]{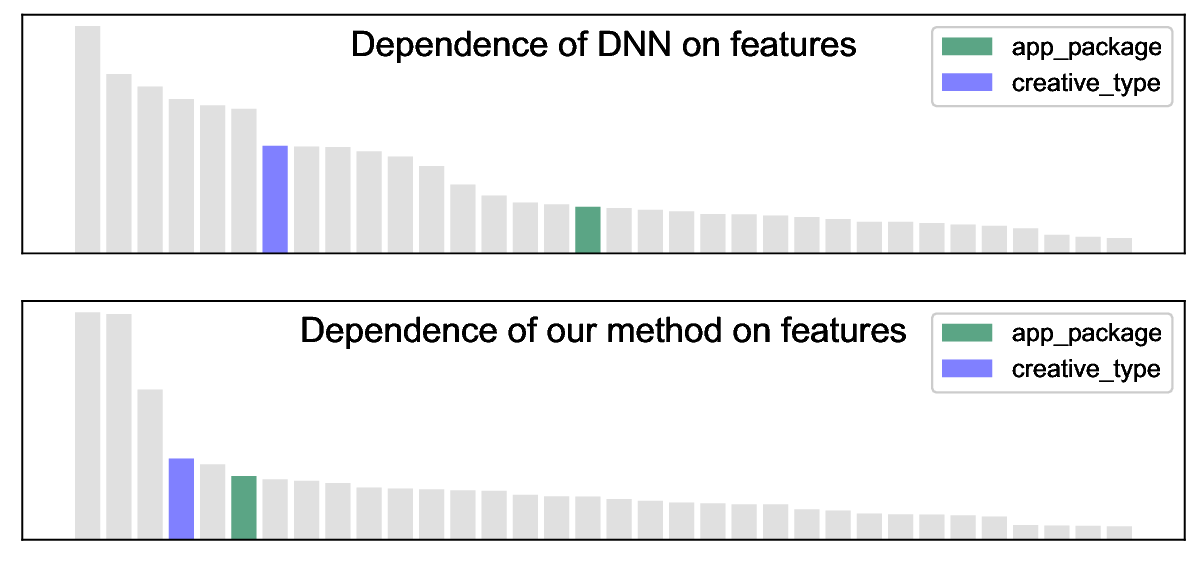}
  \vspace{-0.15in}
  \caption{Our method effectively improves the model's dependence on two target features: \textit{app\_package} and \textit{creative\_type}.}
  \vspace{-0.15in}
  \label{fig.emphasis}
  \Description{}
\end{figure}

In this section, we validate on the industrial dataset that our method can effectively emphasize domain-sensitive features, thus avoiding the undesired neglect shown in Fig.~\ref{fig.neglect}.
Specifically, we select two features, \textit{app\_package} and \textit{creative\_type}, from the top five domain-sensitive features as the targets to emphasize, i.e., the input of the Domain-Sensitive Feature Extractor.
We then calculate our method's dependence on the two target features via Integrated Gradients~\cite{ig_sundararajan2017axiomatic_ig} and compare it with DNN.
Same as Fig.~\ref{fig.neglect}, we sort features according to the model's dependence and highlight the position of target features in Fig.~\ref{fig.emphasis}.
Our method evidently moves the dependence ranking of two target features ahead, indicating a successful emphasis on domain-sensitive features.

\subsection{Accuracy of Domain-Sensitive Feature Attribution (RQ3)}
\begin{figure}[!h]
  \centering
    \includegraphics[width=0.99\linewidth]{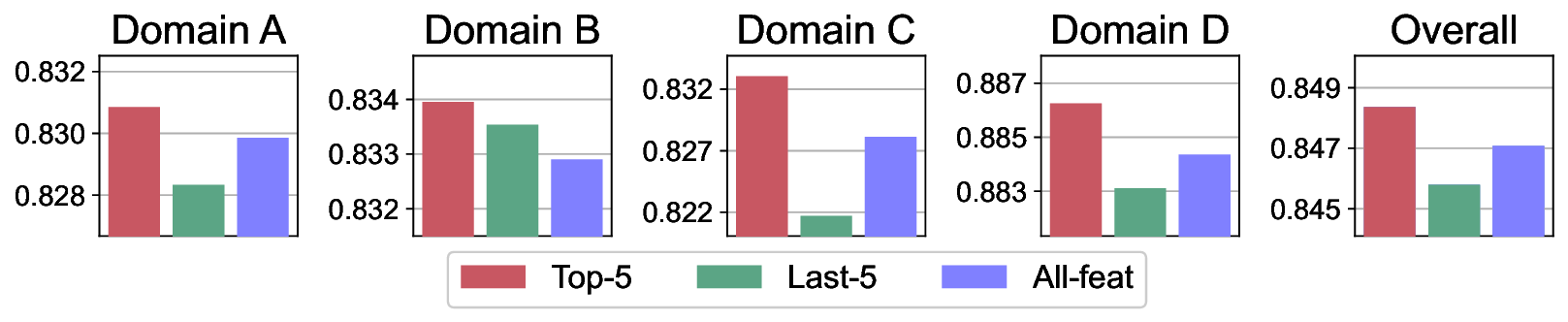}
  \vspace{-0.1in}
  \caption{Results (AUC) of different selections of domain-sensitive features.}
  \vspace{-0.2in}
  \label{fig.ablation_attr}
  \Description{}
\end{figure}

The result of domain-sensitive feature attribution determines the input of the Extractor in our model.
If the input features are not sensitive to the domain, meaning they carry little domain-specific information, the improvement on the model's performance will be limited.
Therefore, we design the following experiment to verify the accuracy of our attribution method.
On the industrial dataset, we conduct the attribution and then separately use the five most domain-sensitive features (Top-5), the five least domain-sensitive features (Last-5), and all features (All-feat) as the input to the Extractor.
Then we compare the model's multi-domain performance in each case. 
As shown in Fig.~\ref{fig.ablation_attr}, Top-5 achieves the best performance across all domains. 
For All-feat, redundant insensitive features may introduce interference and affect the information extraction from domain-sensitive features, leading to a slight decrease in performance.
As expected, Last-5 has a large decrease in performance across nearly all domains.

\subsection{Computational Efficiency Analysis (RQ4)}
\begin{figure}[!h]
  \centering
    \includegraphics[width=0.75\linewidth]{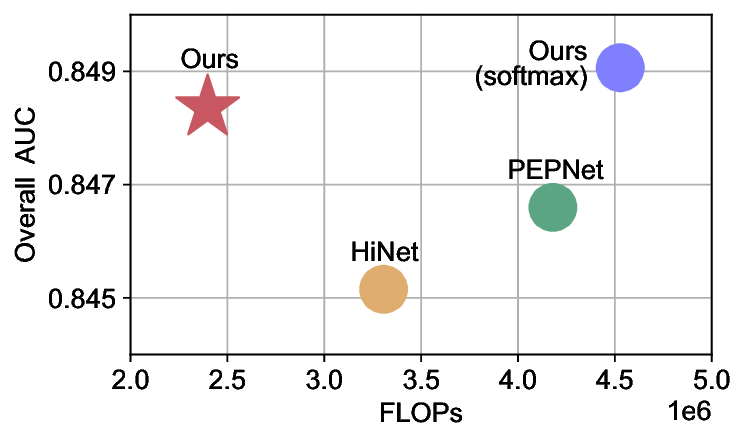}
  \vspace{-0.15in}
  \caption{The performance and computational cost of our method and several primary baselines.}
  \vspace{-0.2in}
  \label{fig.flops}
  \Description{}
\end{figure}

Computational costs typically limit the online deployment of recommendation models.
Therefore, we expect the model to have outstanding predicting performance while maintaining high computational efficiency.
Thanks to the application of linear attentions, our method satisfies this expectation.
In this section, we compare the computational cost of our method with several primary baselines in terms of FLOPs, i.e., the Floating Point Operations.
As shown in Fig.~\ref{fig.flops}, our method not only outperforms the baselines in AUC but also has lower computational cost.
In addition, as a substitute of the softmax attention, linear attentions adopted in the Retriever significantly reduce FLOPs, while maintaining almost the same AUC.
Overall, our method is friendly to online deployment.

\subsection{Online A/B Test (RQ5)}
To validate the effectiveness of our method in real-world recommender systems, we conduct an online A/B test for two consecutive weeks on Huawei's online advertising platform.
The platform provides ad recommendations to tens of millions of active users daily.
The baseline is a highly optimized multi-domain recommendation model.
Our method and the baseline are each allocated 10\% of the online traffic.
Each model is deployed on a single cluster with the same hardware configuration.
The training data of each model is sourced from the latest online logs and processed through identical pipelines.
Over the two-week test, our method significantly improves business metrics compared to the baseline, including an 8.168\% increase in revenue per mille (RPM) and a 2.86\% increase in effective cost per mille (eCPM).
\section{Conclusion}
In this paper, we consider the multi-domain recommendation problem from the perspective of feature distribution.
We refer to features that exhibit significant 
inter-domain differences in feature distributions and effects on model predictions as domain-sensitive features. 
Experiments indicate that existing methods may neglect domain-sensitive features, leading to insufficient learning of domain distinctions. 
To address this issue, we propose a Domain-Sensitive Feature Attribution method to select domain-sensitive features that best represent domain distinctions from the entire feature set.
Furthermore, we design a Domain-Sensitive Feature Memory that extracts domain-specific information from domain-sensitive features for the model to retrieve and utilize, thus enhancing the awareness of domain distinctions.
Comprehensive offline and online experiments demonstrate the effectiveness of our method in emphasizing domain-sensitive features and boosting multi-domain recommendation performance, as well as its decent computational efficiency for online deployment.

\bibliographystyle{ACM-Reference-Format}
\bibliography{main}


\begin{thebibliography}{48}


\ifx \showCODEN    \undefined \def \showCODEN     #1{\unskip}     \fi
\ifx \showDOI      \undefined \def \showDOI       #1{#1}\fi
\ifx \showISBNx    \undefined \def \showISBNx     #1{\unskip}     \fi
\ifx \showISBNxiii \undefined \def \showISBNxiii  #1{\unskip}     \fi
\ifx \showISSN     \undefined \def \showISSN      #1{\unskip}     \fi
\ifx \showLCCN     \undefined \def \showLCCN      #1{\unskip}     \fi
\ifx \shownote     \undefined \def \shownote      #1{#1}          \fi
\ifx \showarticletitle \undefined \def \showarticletitle #1{#1}   \fi
\ifx \showURL      \undefined \def \showURL       {\relax}        \fi
\providecommand\bibfield[2]{#2}
\providecommand\bibinfo[2]{#2}
\providecommand\natexlab[1]{#1}
\providecommand\showeprint[2][]{arXiv:#2}

\bibitem[Arjovsky et~al\mbox{.}(2017)]%
        {wgan_arjovsky2017wasserstein_wgan}
\bibfield{author}{\bibinfo{person}{Martin Arjovsky}, \bibinfo{person}{Soumith Chintala}, {and} \bibinfo{person}{L{\'e}on Bottou}.} \bibinfo{year}{2017}\natexlab{}.
\newblock \showarticletitle{Wasserstein generative adversarial networks}. In \bibinfo{booktitle}{\emph{ICML}}. PMLR, \bibinfo{pages}{214--223}.
\newblock


\bibitem[Bach et~al\mbox{.}(2015)]%
        {lrp_bach2015pixel_lrp}
\bibfield{author}{\bibinfo{person}{Sebastian Bach}, \bibinfo{person}{Alexander Binder}, \bibinfo{person}{Gr{\'e}goire Montavon}, \bibinfo{person}{Frederick Klauschen}, \bibinfo{person}{Klaus-Robert M{\"u}ller}, {and} \bibinfo{person}{Wojciech Samek}.} \bibinfo{year}{2015}\natexlab{}.
\newblock \showarticletitle{On pixel-wise explanations for non-linear classifier decisions by layer-wise relevance propagation}.
\newblock \bibinfo{journal}{\emph{PloS one}} \bibinfo{volume}{10}, \bibinfo{number}{7} (\bibinfo{year}{2015}), \bibinfo{pages}{e0130140}.
\newblock


\bibitem[Caruana(1997)]%
        {sharedbottom_caruana1997multitask}
\bibfield{author}{\bibinfo{person}{Rich Caruana}.} \bibinfo{year}{1997}\natexlab{}.
\newblock \showarticletitle{Multitask learning}.
\newblock \bibinfo{journal}{\emph{Machine learning}}  \bibinfo{volume}{28} (\bibinfo{year}{1997}), \bibinfo{pages}{41--75}.
\newblock


\bibitem[Chang et~al\mbox{.}(2023)]%
        {pepnet_chang2023pepnet}
\bibfield{author}{\bibinfo{person}{Jianxin Chang}, \bibinfo{person}{Chenbin Zhang}, \bibinfo{person}{Yiqun Hui}, \bibinfo{person}{Dewei Leng}, \bibinfo{person}{Yanan Niu}, \bibinfo{person}{Yang Song}, {and} \bibinfo{person}{Kun Gai}.} \bibinfo{year}{2023}\natexlab{}.
\newblock \showarticletitle{Pepnet: Parameter and embedding personalized network for infusing with personalized prior information}. In \bibinfo{booktitle}{\emph{SIGKDD}}. \bibinfo{pages}{3795--3804}.
\newblock


\bibitem[Chapelle et~al\mbox{.}(2014)]%
        {ctr_chapelle2014simple_ctr}
\bibfield{author}{\bibinfo{person}{Olivier Chapelle}, \bibinfo{person}{Eren Manavoglu}, {and} \bibinfo{person}{Romer Rosales}.} \bibinfo{year}{2014}\natexlab{}.
\newblock \showarticletitle{Simple and scalable response prediction for display advertising}.
\newblock \bibinfo{journal}{\emph{TIST}} \bibinfo{volume}{5}, \bibinfo{number}{4} (\bibinfo{year}{2014}), \bibinfo{pages}{1--34}.
\newblock


\bibitem[Chen et~al\mbox{.}(2016)]%
        {ctr_survey_chen2016deep_ctr}
\bibfield{author}{\bibinfo{person}{Junxuan Chen}, \bibinfo{person}{Baigui Sun}, \bibinfo{person}{Hao Li}, \bibinfo{person}{Hongtao Lu}, {and} \bibinfo{person}{Xian-Sheng Hua}.} \bibinfo{year}{2016}\natexlab{}.
\newblock \showarticletitle{Deep ctr prediction in display advertising}. In \bibinfo{booktitle}{\emph{ACM MM}}. \bibinfo{pages}{811--820}.
\newblock


\bibitem[Choromanski et~al\mbox{.}(2020)]%
        {performer_choromanski2020rethinking_performer}
\bibfield{author}{\bibinfo{person}{Krzysztof Choromanski}, \bibinfo{person}{Valerii Likhosherstov}, \bibinfo{person}{David Dohan}, \bibinfo{person}{Xingyou Song}, \bibinfo{person}{Andreea Gane}, \bibinfo{person}{Tamas Sarlos}, \bibinfo{person}{Peter Hawkins}, \bibinfo{person}{Jared Davis}, \bibinfo{person}{Afroz Mohiuddin}, \bibinfo{person}{Lukasz Kaiser}, {et~al\mbox{.}}} \bibinfo{year}{2020}\natexlab{}.
\newblock \showarticletitle{Rethinking attention with performers}.
\newblock \bibinfo{journal}{\emph{arXiv preprint arXiv:2009.14794}} (\bibinfo{year}{2020}).
\newblock


\bibitem[Clevert et~al\mbox{.}(2015)]%
        {elu_clevert2015fast_elu}
\bibfield{author}{\bibinfo{person}{Djork-Arn{\'e} Clevert}, \bibinfo{person}{Thomas Unterthiner}, {and} \bibinfo{person}{Sepp Hochreiter}.} \bibinfo{year}{2015}\natexlab{}.
\newblock \showarticletitle{Fast and accurate deep network learning by exponential linear units (elus)}.
\newblock \bibinfo{journal}{\emph{arXiv preprint arXiv:1511.07289}} (\bibinfo{year}{2015}).
\newblock


\bibitem[Covington et~al\mbox{.}(2016)]%
        {dnn_covington2016deep}
\bibfield{author}{\bibinfo{person}{Paul Covington}, \bibinfo{person}{Jay Adams}, {and} \bibinfo{person}{Emre Sargin}.} \bibinfo{year}{2016}\natexlab{}.
\newblock \showarticletitle{Deep neural networks for youtube recommendations}. In \bibinfo{booktitle}{\emph{RecSys}}. \bibinfo{pages}{191--198}.
\newblock


\bibitem[Deng et~al\mbox{.}(2024)]%
        {14_attr_deng2024unifying_14}
\bibfield{author}{\bibinfo{person}{Huiqi Deng}, \bibinfo{person}{Na Zou}, \bibinfo{person}{Mengnan Du}, \bibinfo{person}{Weifu Chen}, \bibinfo{person}{Guocan Feng}, \bibinfo{person}{Ziwei Yang}, \bibinfo{person}{Zheyang Li}, {and} \bibinfo{person}{Quanshi Zhang}.} \bibinfo{year}{2024}\natexlab{}.
\newblock \showarticletitle{Unifying Fourteen Post-hoc Attribution Methods with Taylor Interactions}.
\newblock \bibinfo{journal}{\emph{TPAMI}} (\bibinfo{year}{2024}).
\newblock


\bibitem[Flamary et~al\mbox{.}(2021)]%
        {pot_flamary2021pot}
\bibfield{author}{\bibinfo{person}{R{\'e}mi Flamary}, \bibinfo{person}{Nicolas Courty}, \bibinfo{person}{Alexandre Gramfort}, \bibinfo{person}{Mokhtar~Z. Alaya}, \bibinfo{person}{Aur{\'e}lie Boisbunon}, \bibinfo{person}{Stanislas Chambon}, \bibinfo{person}{Laetitia Chapel}, \bibinfo{person}{Adrien Corenflos}, \bibinfo{person}{Kilian Fatras}, \bibinfo{person}{Nemo Fournier}, \bibinfo{person}{L{\'e}o Gautheron}, \bibinfo{person}{Nathalie~T.H. Gayraud}, \bibinfo{person}{Hicham Janati}, \bibinfo{person}{Alain Rakotomamonjy}, \bibinfo{person}{Ievgen Redko}, \bibinfo{person}{Antoine Rolet}, \bibinfo{person}{Antony Schutz}, \bibinfo{person}{Vivien Seguy}, \bibinfo{person}{Danica~J. Sutherland}, \bibinfo{person}{Romain Tavenard}, \bibinfo{person}{Alexander Tong}, {and} \bibinfo{person}{Titouan Vayer}.} \bibinfo{year}{2021}\natexlab{}.
\newblock \showarticletitle{POT: Python Optimal Transport}.
\newblock \bibinfo{journal}{\emph{JMLR}} \bibinfo{volume}{22}, \bibinfo{number}{78} (\bibinfo{year}{2021}), \bibinfo{pages}{1--8}.
\newblock
\urldef\tempurl%
\url{http://jmlr.org/papers/v22/20-451.html}
\showURL{%
\tempurl}


\bibitem[Guidotti et~al\mbox{.}(2018)]%
        {attr_survey_guidotti2018survey}
\bibfield{author}{\bibinfo{person}{Riccardo Guidotti}, \bibinfo{person}{Anna Monreale}, \bibinfo{person}{Salvatore Ruggieri}, \bibinfo{person}{Franco Turini}, \bibinfo{person}{Fosca Giannotti}, {and} \bibinfo{person}{Dino Pedreschi}.} \bibinfo{year}{2018}\natexlab{}.
\newblock \showarticletitle{A survey of methods for explaining black box models}.
\newblock \bibinfo{journal}{\emph{ACM computing surveys (CSUR)}} \bibinfo{volume}{51}, \bibinfo{number}{5} (\bibinfo{year}{2018}), \bibinfo{pages}{1--42}.
\newblock


\bibitem[He et~al\mbox{.}(2020)]%
        {dadnn_he2020dadnn}
\bibfield{author}{\bibinfo{person}{Junyou He}, \bibinfo{person}{Guibao Mei}, \bibinfo{person}{Feng Xing}, \bibinfo{person}{Xiaorui Yang}, \bibinfo{person}{Yongjun Bao}, {and} \bibinfo{person}{Weipeng Yan}.} \bibinfo{year}{2020}\natexlab{}.
\newblock \showarticletitle{Dadnn: multi-scene ctr prediction via domain-aware deep neural network}.
\newblock \bibinfo{journal}{\emph{arXiv preprint arXiv:2011.11938}} (\bibinfo{year}{2020}).
\newblock


\bibitem[He et~al\mbox{.}(2016)]%
        {residual_he2016deep_residual}
\bibfield{author}{\bibinfo{person}{Kaiming He}, \bibinfo{person}{Xiangyu Zhang}, \bibinfo{person}{Shaoqing Ren}, {and} \bibinfo{person}{Jian Sun}.} \bibinfo{year}{2016}\natexlab{}.
\newblock \showarticletitle{Deep residual learning for image recognition}. In \bibinfo{booktitle}{\emph{CVPR}}. \bibinfo{pages}{770--778}.
\newblock


\bibitem[Hospedales et~al\mbox{.}(2021)]%
        {meta_hospedales2021meta}
\bibfield{author}{\bibinfo{person}{Timothy Hospedales}, \bibinfo{person}{Antreas Antoniou}, \bibinfo{person}{Paul Micaelli}, {and} \bibinfo{person}{Amos Storkey}.} \bibinfo{year}{2021}\natexlab{}.
\newblock \showarticletitle{Meta-learning in neural networks: A survey}.
\newblock \bibinfo{journal}{\emph{TPAMI}} \bibinfo{volume}{44}, \bibinfo{number}{9} (\bibinfo{year}{2021}), \bibinfo{pages}{5149--5169}.
\newblock


\bibitem[Kantorovich(1960)]%
        {wdis_kantorovich1960mathematical_wdis}
\bibfield{author}{\bibinfo{person}{Leonid~V Kantorovich}.} \bibinfo{year}{1960}\natexlab{}.
\newblock \showarticletitle{Mathematical methods of organizing and planning production}.
\newblock \bibinfo{journal}{\emph{Management science}} \bibinfo{volume}{6}, \bibinfo{number}{4} (\bibinfo{year}{1960}), \bibinfo{pages}{366--422}.
\newblock


\bibitem[Katharopoulos et~al\mbox{.}(2020)]%
        {linear_attn_katharopoulos2020transformers_linear_attention}
\bibfield{author}{\bibinfo{person}{Angelos Katharopoulos}, \bibinfo{person}{Apoorv Vyas}, \bibinfo{person}{Nikolaos Pappas}, {and} \bibinfo{person}{Fran{\c{c}}ois Fleuret}.} \bibinfo{year}{2020}\natexlab{}.
\newblock \showarticletitle{Transformers are rnns: Fast autoregressive transformers with linear attention}. In \bibinfo{booktitle}{\emph{ICML}}. PMLR, \bibinfo{pages}{5156--5165}.
\newblock


\bibitem[Li et~al\mbox{.}(2020)]%
        {hmoe_li2020improving}
\bibfield{author}{\bibinfo{person}{Pengcheng Li}, \bibinfo{person}{Runze Li}, \bibinfo{person}{Qing Da}, \bibinfo{person}{An-Xiang Zeng}, {and} \bibinfo{person}{Lijun Zhang}.} \bibinfo{year}{2020}\natexlab{}.
\newblock \showarticletitle{Improving multi-scenario learning to rank in e-commerce by exploiting task relationships in the label space}. In \bibinfo{booktitle}{\emph{CIKM}}. \bibinfo{pages}{2605--2612}.
\newblock


\bibitem[Li et~al\mbox{.}(2023)]%
        {hamur_li2023hamur}
\bibfield{author}{\bibinfo{person}{Xiaopeng Li}, \bibinfo{person}{Fan Yan}, \bibinfo{person}{Xiangyu Zhao}, \bibinfo{person}{Yichao Wang}, \bibinfo{person}{Bo Chen}, \bibinfo{person}{Huifeng Guo}, {and} \bibinfo{person}{Ruiming Tang}.} \bibinfo{year}{2023}\natexlab{}.
\newblock \showarticletitle{Hamur: Hyper adapter for multi-domain recommendation}. In \bibinfo{booktitle}{\emph{CIKM}}. \bibinfo{pages}{1268--1277}.
\newblock


\bibitem[Lundberg and Lee(2017)]%
        {deepshap_lundberg2017unified_deepshap}
\bibfield{author}{\bibinfo{person}{Scott~M Lundberg} {and} \bibinfo{person}{Su-In Lee}.} \bibinfo{year}{2017}\natexlab{}.
\newblock \showarticletitle{A unified approach to interpreting model predictions}.
\newblock \bibinfo{journal}{\emph{NIPS}}  \bibinfo{volume}{30} (\bibinfo{year}{2017}).
\newblock


\bibitem[Ma et~al\mbox{.}(2018)]%
        {mmoe_ma2018modeling_mmoe}
\bibfield{author}{\bibinfo{person}{Jiaqi Ma}, \bibinfo{person}{Zhe Zhao}, \bibinfo{person}{Xinyang Yi}, \bibinfo{person}{Jilin Chen}, \bibinfo{person}{Lichan Hong}, {and} \bibinfo{person}{Ed~H Chi}.} \bibinfo{year}{2018}\natexlab{}.
\newblock \showarticletitle{Modeling task relationships in multi-task learning with multi-gate mixture-of-experts}. In \bibinfo{booktitle}{\emph{SIGKDD}}. \bibinfo{pages}{1930--1939}.
\newblock


\bibitem[Min et~al\mbox{.}(2023)]%
        {satrans_min2023scenario_satrans}
\bibfield{author}{\bibinfo{person}{Erxue Min}, \bibinfo{person}{Da Luo}, \bibinfo{person}{Kangyi Lin}, \bibinfo{person}{Chunzhen Huang}, {and} \bibinfo{person}{Yang Liu}.} \bibinfo{year}{2023}\natexlab{}.
\newblock \showarticletitle{Scenario-Adaptive Feature Interaction for Click-Through Rate Prediction}. In \bibinfo{booktitle}{\emph{SIGKDD}}. \bibinfo{pages}{4661--4672}.
\newblock


\bibitem[Richardson et~al\mbox{.}(2007)]%
        {ctr_richardson2007predicting_ctr}
\bibfield{author}{\bibinfo{person}{Matthew Richardson}, \bibinfo{person}{Ewa Dominowska}, {and} \bibinfo{person}{Robert Ragno}.} \bibinfo{year}{2007}\natexlab{}.
\newblock \showarticletitle{Predicting clicks: estimating the click-through rate for new ads}. In \bibinfo{booktitle}{\emph{WWW}}. \bibinfo{pages}{521--530}.
\newblock


\bibitem[Selvaraju et~al\mbox{.}(2017)]%
        {grad_cam_selvaraju2017grad_cam}
\bibfield{author}{\bibinfo{person}{Ramprasaath~R Selvaraju}, \bibinfo{person}{Michael Cogswell}, \bibinfo{person}{Abhishek Das}, \bibinfo{person}{Ramakrishna Vedantam}, \bibinfo{person}{Devi Parikh}, {and} \bibinfo{person}{Dhruv Batra}.} \bibinfo{year}{2017}\natexlab{}.
\newblock \showarticletitle{Grad-cam: Visual explanations from deep networks via gradient-based localization}. In \bibinfo{booktitle}{\emph{ICCV}}. \bibinfo{pages}{618--626}.
\newblock


\bibitem[Shen et~al\mbox{.}(2021a)]%
        {sarnet_shen2021sar_net}
\bibfield{author}{\bibinfo{person}{Qijie Shen}, \bibinfo{person}{Wanjie Tao}, \bibinfo{person}{Jing Zhang}, \bibinfo{person}{Hong Wen}, \bibinfo{person}{Zulong Chen}, {and} \bibinfo{person}{Quan Lu}.} \bibinfo{year}{2021}\natexlab{a}.
\newblock \showarticletitle{Sar-net: a scenario-aware ranking network for personalized fair recommendation in hundreds of travel scenarios}. In \bibinfo{booktitle}{\emph{CIKM}}. \bibinfo{pages}{4094--4103}.
\newblock


\bibitem[Shen et~al\mbox{.}(2021b)]%
        {efficien_attention_shen2021efficien_attention}
\bibfield{author}{\bibinfo{person}{Zhuoran Shen}, \bibinfo{person}{Mingyuan Zhang}, \bibinfo{person}{Haiyu Zhao}, \bibinfo{person}{Shuai Yi}, {and} \bibinfo{person}{Hongsheng Li}.} \bibinfo{year}{2021}\natexlab{b}.
\newblock \showarticletitle{Efficient attention: Attention with linear complexities}. In \bibinfo{booktitle}{\emph{WACV}}. \bibinfo{pages}{3531--3539}.
\newblock


\bibitem[Sheng et~al\mbox{.}(2021)]%
        {star_sheng2021one_star}
\bibfield{author}{\bibinfo{person}{Xiang-Rong Sheng}, \bibinfo{person}{Liqin Zhao}, \bibinfo{person}{Guorui Zhou}, \bibinfo{person}{Xinyao Ding}, \bibinfo{person}{Binding Dai}, \bibinfo{person}{Qiang Luo}, \bibinfo{person}{Siran Yang}, \bibinfo{person}{Jingshan Lv}, \bibinfo{person}{Chi Zhang}, \bibinfo{person}{Hongbo Deng}, {et~al\mbox{.}}} \bibinfo{year}{2021}\natexlab{}.
\newblock \showarticletitle{One model to serve all: Star topology adaptive recommender for multi-domain ctr prediction}. In \bibinfo{booktitle}{\emph{CIKM}}. \bibinfo{pages}{4104--4113}.
\newblock


\bibitem[Shrikumar et~al\mbox{.}(2017)]%
        {deeplift_shrikumar2017learning_deeplift}
\bibfield{author}{\bibinfo{person}{Avanti Shrikumar}, \bibinfo{person}{Peyton Greenside}, {and} \bibinfo{person}{Anshul Kundaje}.} \bibinfo{year}{2017}\natexlab{}.
\newblock \showarticletitle{Learning important features through propagating activation differences}. In \bibinfo{booktitle}{\emph{ICML}}. PMLR, \bibinfo{pages}{3145--3153}.
\newblock


\bibitem[Sundararajan et~al\mbox{.}(2017)]%
        {ig_sundararajan2017axiomatic_ig}
\bibfield{author}{\bibinfo{person}{Mukund Sundararajan}, \bibinfo{person}{Ankur Taly}, {and} \bibinfo{person}{Qiqi Yan}.} \bibinfo{year}{2017}\natexlab{}.
\newblock \showarticletitle{Axiomatic attribution for deep networks}. In \bibinfo{booktitle}{\emph{ICML}}. PMLR, \bibinfo{pages}{3319--3328}.
\newblock


\bibitem[Swietojanski et~al\mbox{.}(2016)]%
        {lhuc_swietojanski2016learning}
\bibfield{author}{\bibinfo{person}{Pawel Swietojanski}, \bibinfo{person}{Jinyu Li}, {and} \bibinfo{person}{Steve Renals}.} \bibinfo{year}{2016}\natexlab{}.
\newblock \showarticletitle{Learning hidden unit contributions for unsupervised acoustic model adaptation}.
\newblock \bibinfo{journal}{\emph{IEEE/ACM Transactions on Audio, Speech, and Language Processing}} \bibinfo{volume}{24}, \bibinfo{number}{8} (\bibinfo{year}{2016}), \bibinfo{pages}{1450--1463}.
\newblock


\bibitem[Tanaka et~al\mbox{.}(2020)]%
        {synflow_tanaka2020pruning_synflow}
\bibfield{author}{\bibinfo{person}{Hidenori Tanaka}, \bibinfo{person}{Daniel Kunin}, \bibinfo{person}{Daniel~L Yamins}, {and} \bibinfo{person}{Surya Ganguli}.} \bibinfo{year}{2020}\natexlab{}.
\newblock \showarticletitle{Pruning neural networks without any data by iteratively conserving synaptic flow}.
\newblock \bibinfo{journal}{\emph{NIPS}}  \bibinfo{volume}{33} (\bibinfo{year}{2020}), \bibinfo{pages}{6377--6389}.
\newblock


\bibitem[Tang et~al\mbox{.}(2020)]%
        {ple_tang2020progressive_ple}
\bibfield{author}{\bibinfo{person}{Hongyan Tang}, \bibinfo{person}{Junning Liu}, \bibinfo{person}{Ming Zhao}, {and} \bibinfo{person}{Xudong Gong}.} \bibinfo{year}{2020}\natexlab{}.
\newblock \showarticletitle{Progressive layered extraction (ple): A novel multi-task learning (mtl) model for personalized recommendations}. In \bibinfo{booktitle}{\emph{RecSys}}. \bibinfo{pages}{269--278}.
\newblock


\bibitem[Tian et~al\mbox{.}(2023)]%
        {maria_tian2023multi}
\bibfield{author}{\bibinfo{person}{Yu Tian}, \bibinfo{person}{Bofang Li}, \bibinfo{person}{Si Chen}, \bibinfo{person}{Xubin Li}, \bibinfo{person}{Hongbo Deng}, \bibinfo{person}{Jian Xu}, \bibinfo{person}{Bo Zheng}, \bibinfo{person}{Qian Wang}, {and} \bibinfo{person}{Chenliang Li}.} \bibinfo{year}{2023}\natexlab{}.
\newblock \showarticletitle{Multi-Scenario Ranking with Adaptive Feature Learning}. In \bibinfo{booktitle}{\emph{SIGIR}}. \bibinfo{pages}{517--526}.
\newblock


\bibitem[Vaswani et~al\mbox{.}(2017)]%
        {attention_vaswani2017attention}
\bibfield{author}{\bibinfo{person}{Ashish Vaswani}, \bibinfo{person}{Noam Shazeer}, \bibinfo{person}{Niki Parmar}, \bibinfo{person}{Jakob Uszkoreit}, \bibinfo{person}{Llion Jones}, \bibinfo{person}{Aidan~N Gomez}, \bibinfo{person}{{\L}ukasz Kaiser}, {and} \bibinfo{person}{Illia Polosukhin}.} \bibinfo{year}{2017}\natexlab{}.
\newblock \showarticletitle{Attention is all you need}.
\newblock \bibinfo{journal}{\emph{NIPS}}  \bibinfo{volume}{30} (\bibinfo{year}{2017}).
\newblock


\bibitem[Wang et~al\mbox{.}(2020)]%
        {linformer_wang2020linformer}
\bibfield{author}{\bibinfo{person}{Sinong Wang}, \bibinfo{person}{Belinda~Z Li}, \bibinfo{person}{Madian Khabsa}, \bibinfo{person}{Han Fang}, {and} \bibinfo{person}{Hao Ma}.} \bibinfo{year}{2020}\natexlab{}.
\newblock \showarticletitle{Linformer: Self-attention with linear complexity}.
\newblock \bibinfo{journal}{\emph{arXiv preprint arXiv:2006.04768}} (\bibinfo{year}{2020}).
\newblock


\bibitem[Wang et~al\mbox{.}(2023)]%
        {multisfs_wang2023single}
\bibfield{author}{\bibinfo{person}{Yejing Wang}, \bibinfo{person}{Zhaocheng Du}, \bibinfo{person}{Xiangyu Zhao}, \bibinfo{person}{Bo Chen}, \bibinfo{person}{Huifeng Guo}, \bibinfo{person}{Ruiming Tang}, {and} \bibinfo{person}{Zhenhua Dong}.} \bibinfo{year}{2023}\natexlab{}.
\newblock \showarticletitle{Single-shot Feature Selection for Multi-task Recommendations}. In \bibinfo{booktitle}{\emph{SIGIR}}. \bibinfo{pages}{341--351}.
\newblock


\bibitem[Wang et~al\mbox{.}(2022)]%
        {autofield_wang2022autofield}
\bibfield{author}{\bibinfo{person}{Yejing Wang}, \bibinfo{person}{Xiangyu Zhao}, \bibinfo{person}{Tong Xu}, {and} \bibinfo{person}{Xian Wu}.} \bibinfo{year}{2022}\natexlab{}.
\newblock \showarticletitle{Autofield: Automating feature selection in deep recommender systems}. In \bibinfo{booktitle}{\emph{WWW}}. \bibinfo{pages}{1977--1986}.
\newblock


\bibitem[Xu et~al\mbox{.}(2023)]%
        {musenet_xu2023musenet}
\bibfield{author}{\bibinfo{person}{Senrong Xu}, \bibinfo{person}{Liangyue Li}, \bibinfo{person}{Yuan Yao}, \bibinfo{person}{Zulong Chen}, \bibinfo{person}{Han Wu}, \bibinfo{person}{Quan Lu}, {and} \bibinfo{person}{Hanghang Tong}.} \bibinfo{year}{2023}\natexlab{}.
\newblock \showarticletitle{MUSENET: Multi-scenario learning for repeat-aware personalized recommendation}. In \bibinfo{booktitle}{\emph{WSDM}}. \bibinfo{pages}{517--525}.
\newblock


\bibitem[Yan et~al\mbox{.}(2022)]%
        {apg_yan2022apg}
\bibfield{author}{\bibinfo{person}{Bencheng Yan}, \bibinfo{person}{Pengjie Wang}, \bibinfo{person}{Kai Zhang}, \bibinfo{person}{Feng Li}, \bibinfo{person}{Hongbo Deng}, \bibinfo{person}{Jian Xu}, {and} \bibinfo{person}{Bo Zheng}.} \bibinfo{year}{2022}\natexlab{}.
\newblock \showarticletitle{Apg: Adaptive parameter generation network for click-through rate prediction}.
\newblock \bibinfo{journal}{\emph{NIPS}}  \bibinfo{volume}{35} (\bibinfo{year}{2022}), \bibinfo{pages}{24740--24752}.
\newblock


\bibitem[Yang et~al\mbox{.}(2022)]%
        {adasparse_yang2022adasparse}
\bibfield{author}{\bibinfo{person}{Xuanhua Yang}, \bibinfo{person}{Xiaoyu Peng}, \bibinfo{person}{Penghui Wei}, \bibinfo{person}{Shaoguo Liu}, \bibinfo{person}{Liang Wang}, {and} \bibinfo{person}{Bo Zheng}.} \bibinfo{year}{2022}\natexlab{}.
\newblock \showarticletitle{Adasparse: Learning adaptively sparse structures for multi-domain click-through rate prediction}. In \bibinfo{booktitle}{\emph{CIKM}}. \bibinfo{pages}{4635--4639}.
\newblock


\bibitem[Zeiler and Fergus(2014)]%
        {occlusion_1_zeiler2014visualizing_occlusion_1}
\bibfield{author}{\bibinfo{person}{Matthew~D Zeiler} {and} \bibinfo{person}{Rob Fergus}.} \bibinfo{year}{2014}\natexlab{}.
\newblock \showarticletitle{Visualizing and understanding convolutional networks}. In \bibinfo{booktitle}{\emph{ECCV}}. Springer, \bibinfo{pages}{818--833}.
\newblock


\bibitem[Zhang et~al\mbox{.}(2022)]%
        {m2m_zhang2022leaving_m2m}
\bibfield{author}{\bibinfo{person}{Qianqian Zhang}, \bibinfo{person}{Xinru Liao}, \bibinfo{person}{Quan Liu}, \bibinfo{person}{Jian Xu}, {and} \bibinfo{person}{Bo Zheng}.} \bibinfo{year}{2022}\natexlab{}.
\newblock \showarticletitle{Leaving no one behind: A multi-scenario multi-task meta learning approach for advertiser modeling}. In \bibinfo{booktitle}{\emph{WSDM}}. \bibinfo{pages}{1368--1376}.
\newblock


\bibitem[Zhang et~al\mbox{.}(2021)]%
        {attr_survey_zhang2021survey}
\bibfield{author}{\bibinfo{person}{Yu Zhang}, \bibinfo{person}{Peter Ti{\v{n}}o}, \bibinfo{person}{Ale{\v{s}} Leonardis}, {and} \bibinfo{person}{Ke Tang}.} \bibinfo{year}{2021}\natexlab{}.
\newblock \showarticletitle{A survey on neural network interpretability}.
\newblock \bibinfo{journal}{\emph{IEEE Transactions on Emerging Topics in Computational Intelligence}} \bibinfo{volume}{5}, \bibinfo{number}{5} (\bibinfo{year}{2021}), \bibinfo{pages}{726--742}.
\newblock


\bibitem[Zhou et~al\mbox{.}(2019)]%
        {dien_zhou2019deep}
\bibfield{author}{\bibinfo{person}{Guorui Zhou}, \bibinfo{person}{Na Mou}, \bibinfo{person}{Ying Fan}, \bibinfo{person}{Qi Pi}, \bibinfo{person}{Weijie Bian}, \bibinfo{person}{Chang Zhou}, \bibinfo{person}{Xiaoqiang Zhu}, {and} \bibinfo{person}{Kun Gai}.} \bibinfo{year}{2019}\natexlab{}.
\newblock \showarticletitle{Deep interest evolution network for click-through rate prediction}. In \bibinfo{booktitle}{\emph{AAAI}}, Vol.~\bibinfo{volume}{33}. \bibinfo{pages}{5941--5948}.
\newblock


\bibitem[Zhou et~al\mbox{.}(2018)]%
        {din_zhou2018deep}
\bibfield{author}{\bibinfo{person}{Guorui Zhou}, \bibinfo{person}{Xiaoqiang Zhu}, \bibinfo{person}{Chenru Song}, \bibinfo{person}{Ying Fan}, \bibinfo{person}{Han Zhu}, \bibinfo{person}{Xiao Ma}, \bibinfo{person}{Yanghui Yan}, \bibinfo{person}{Junqi Jin}, \bibinfo{person}{Han Li}, {and} \bibinfo{person}{Kun Gai}.} \bibinfo{year}{2018}\natexlab{}.
\newblock \showarticletitle{Deep interest network for click-through rate prediction}. In \bibinfo{booktitle}{\emph{SIGKDD}}. \bibinfo{pages}{1059--1068}.
\newblock


\bibitem[Zhou et~al\mbox{.}(2023)]%
        {hinet_zhou2023hinet}
\bibfield{author}{\bibinfo{person}{Jie Zhou}, \bibinfo{person}{Xianshuai Cao}, \bibinfo{person}{Wenhao Li}, \bibinfo{person}{Lin Bo}, \bibinfo{person}{Kun Zhang}, \bibinfo{person}{Chuan Luo}, {and} \bibinfo{person}{Qian Yu}.} \bibinfo{year}{2023}\natexlab{}.
\newblock \showarticletitle{Hinet: Novel multi-scenario \& multi-task learning with hierarchical information extraction}. In \bibinfo{booktitle}{\emph{ICDE}}. IEEE, \bibinfo{pages}{2969--2975}.
\newblock


\bibitem[Zhu et~al\mbox{.}(2017)]%
        {multidomain_zhu2017optimized_multidomain}
\bibfield{author}{\bibinfo{person}{Han Zhu}, \bibinfo{person}{Junqi Jin}, \bibinfo{person}{Chang Tan}, \bibinfo{person}{Fei Pan}, \bibinfo{person}{Yifan Zeng}, \bibinfo{person}{Han Li}, {and} \bibinfo{person}{Kun Gai}.} \bibinfo{year}{2017}\natexlab{}.
\newblock \showarticletitle{Optimized cost per click in taobao display advertising}. In \bibinfo{booktitle}{\emph{SIGKDD}}. \bibinfo{pages}{2191--2200}.
\newblock


\bibitem[Zintgraf et~al\mbox{.}(2017)]%
        {occlusion_patch_zintgraf2017visualizing_occlusion_patch}
\bibfield{author}{\bibinfo{person}{Luisa~M Zintgraf}, \bibinfo{person}{Taco~S Cohen}, \bibinfo{person}{Tameem Adel}, {and} \bibinfo{person}{Max Welling}.} \bibinfo{year}{2017}\natexlab{}.
\newblock \showarticletitle{Visualizing deep neural network decisions: Prediction difference analysis}.
\newblock \bibinfo{journal}{\emph{arXiv preprint arXiv:1702.04595}} (\bibinfo{year}{2017}).
\newblock


\end{thebibliography}

\end{document}